\documentclass[12pt,a4paper]{article}
\usepackage{amsmath}
\usepackage{graphicx}
\usepackage[dvips]{color}
\begin{document}


\title{
\color{red}
Ward identities for charge and heat currents of particle-particle and particle-hole pairs
}

\color{black}

\author{
Osamu Narikiyo\footnote{
Department of Physics, 
Kyushu University, 
Fukuoka 812-8581, 
Japan; E-mail: narikiyo@phys.kyushu-u.ac.jp}
}

\maketitle
\begin{abstract}
The Ward identities for the charge and heat currents are derived 
for particle-particle and particle-hole pairs. 
They are the exact constraints on the current-vertex functions 
imposed by conservation laws and 
should be satisfied by consistent theories. 
While the Ward identity for the charge current of electrons 
is well established, 
that for the heat current is not understood correctly. 
Thus the correct interpretation is presented. 
On this firm basis the Ward identities for pairs are discussed. 
As the application of the identity 
we \color{red} criticize \color{black} some inconsistent results 
in the studies of 
the superconducting fluctuation transport and 
the transport anomaly in the normal state of high-$T_c$ superconductors. 

PACS 74.25.fc - Electric and thermal conductivity 

PACS 74.40.-n - Fluctuation phenomena 

PACS 05.60.Gg - Quantum transport 
\end{abstract}

\vskip 30pt 

{\bf Introduction.} The Ward identity plays crucial roles 
in various aspects of theoretical physics. 
It is a consequence of a conservation law 
and a basic relation which should be satisfied by consistent theories. 
One of the most effective applications of the Ward identity 
in condensed matter physics is the gauge invariant formulation 
of the Meissner effect in superconductors~\cite{Nam}. 
At the same time 
it leads to the discovery of the Nambu-Goldstone mode 
which appears in the state with spontaneously broken symmetry. 

In this Letter 
two kinds of Ward identity are discussed 
in the context of condensed matter physics. 
One is for the charge current and the other is for the heat current. 
The Ward identity for the charge-current vertex of electrons 
is well-known by the textbook discussion~\cite{Sch}. 
On the other hand, 
there is no literature which summarizes the correct understanding 
of the heat-current vertex. 
Thus we give a summary on the heat-current vertex 
including our original finding. 

The above description concerns the vertex function for electrons. 
However, the main purpose of this Letter is 
to establish the constraint on the vertex function, the Ward identity, 
for particle-particle and particle-hole pairs. 
Although the Ward identity for the charge current 
carried by particle-particle pairs has been discussed 
in the study of superconducting fluctuation transport~\cite{Tsu,AHL}, 
the proof of the identity has not been given. 
Although it is obvious 
that particle-hole pairs, which are charge-neutral, do not carry charge, 
only perturbational results have been reported~\cite{Tak,LM} 
but the rigorous proof of it, 
which can be achieved on the basis of the Ward identity, is absent. 
For the heat current 
there is no literature discussing the Ward identity for pairs. 
These absent discussions are given in this Letter. 

By establishing the rigorous constraint on the vertex function 
we \color{red} criticize \color{black} some inconsistent arguments 
seen in the published results. 
Several examples are discussed as the applications of the Ward identity. 

Detailed calculations are given in the notes at arXiv: 
[N1] $\equiv$ 1108.0815, 
[N2] $\equiv$ 1108.5272, 
[N3] $\equiv$ 1109.1404, 
[N4] $\equiv$ 1112.1513, 
[N5] $\equiv$ 1212.6484, and 
[N6] $\equiv$ 1309.4257. 

{\bf Algebraic proof.} First we show the proof~\cite{Sch} of the Ward identity 
for the charge current of electrons at zero temperature. 
Let us start from the charge conservation law 
\begin{equation}
\nabla \cdot {\bf j}({\bf r},t) + 
{\partial \over \partial t} \rho({\bf r},t) = 0, 
\end{equation}
where $\rho({\bf r},t)$ and ${\bf j}({\bf r},t)$ are 
the charge and current densities at the position ${\bf r}$ and the time $t$. 
The three-current ${\bf j}$ is represented as ${\bf j}=(j_1,j_2,j_3)$. 
In Fourier transformed variables the conservation law becomes 
\begin{equation}
{\bf q} \cdot {\bf j}({\bf q},\omega) - \omega \rho({\bf q},\omega) = 0. 
\end{equation}
By introducing the four-current $j=(j_1,j_2,j_3,j_0)$ with $j_0 \equiv \rho$ 
and the four-vectors $x=(x_1,x_2,x_3,x_0)=({\bf r},t)$ and 
$q=(q_1,q_2,q_3,q_0)=({\bf q},-\omega)$, 
the conservation law is expressed as the vanishing four-divergence, 
\begin{equation}
\sum_{\mu=0}^{3} {\partial \over \partial x_\mu} j_\mu(x) = 0 = 
\sum_{\mu=0}^{3} q_\mu j_\mu(q). 
\label{4div} 
\end{equation}

Here we introduce the three-point function $\Lambda_\mu$ defined by 
\begin{equation}
\Lambda_\mu(x,y,z) = \big\langle T 
\big\{ j_\mu(z) \psi_\uparrow(x) \psi_\uparrow^\dag(y) \big\} \big\rangle, 
\label{Lambda(x,y,z)} 
\end{equation}
where $x$, $y$ and $z$ are the four-vectors in real space, 
$\langle A \rangle$ represents the expectation value of $A$ 
in the ground state, $T$ is the time-ordering operator, and 
$\psi_\uparrow(x)$ and $\psi_\uparrow^\dag(y)$ are 
annihilation and creation operators of $\uparrow$-spin electron. 
Under the conservation law (\ref{4div}) 
the four-divergence of $\Lambda_\mu$ reduces to 
\begin{equation}
\big\langle T 
\big\{ \big[ \rho(z), \psi_\uparrow(x) \big] 
\psi_\uparrow^\dag(y) \big\} \big\rangle 
\delta(z_0-x_0) 
+ 
\big\langle T 
\big\{ \psi_\uparrow(x) 
\big[ \rho(z), \psi_\uparrow^\dag(y) \big] \big\} \big\rangle 
\delta(z_0-y_0). 
\end{equation}
The commutation relations reduce to 
the annihilation and creation of electron charge as 
\begin{equation}
\big[ \rho(z), \psi_\uparrow(x) \big] \delta(z_0-x_0) 
= - e \psi_\uparrow(x) \delta^4(z-x), 
\label{com-ann} 
\end{equation}
and 
\begin{equation}
\big[ \rho(z), \psi_\uparrow^\dag(y) \big] \delta(z_0-y_0) 
= e \psi_\uparrow^\dag(y) \delta^4(z-y), 
\label{com-cre} 
\end{equation}
so that we obtain 
\begin{equation}
\sum_{\mu=0}^3 {\partial \over \partial z_\mu} \Lambda_\mu(x,y,z) = 
- i e G(x,y) \delta^4(z-x) + i e G(x,y) \delta^4(z-y), 
\label{divLambda} 
\end{equation}
where the electron propagator $G(x,y)$  is introduced as 
\begin{equation}
G(x,y) = 
- i \big\langle T \big\{ \psi_\uparrow(x) 
\psi_\uparrow^\dag(y) \big\} \big\rangle. 
\end{equation}
In Fourier-transformed variables (\ref{divLambda}) is expressed as 
\begin{equation}
\sum_{\mu=0}^3 k_\mu \Lambda_\mu(p,k)
= e G(p) - e G(p+k). 
\end{equation}
Since the current vertex $\Gamma_\mu$ is related to $\Lambda_\mu$ as 
\begin{equation}
\Lambda_\mu(p,k) = i G(p) \cdot \Gamma_\mu(p,k) \cdot i G(p+k), 
\end{equation}
we obtain the Ward identity for the charge current [N1,\ N4] as   
\begin{equation}
\sum_{\mu=0}^3 k_\mu \Gamma_\mu(p,k) = 
e G(p)^{-1}- e G(p+k)^{-1}. 
\label{Ward-e} 
\end{equation}
In Fig. 1 (left) the three-point function $ \Lambda_\mu(p,k)$ is depicted 
where the circle represents $\Gamma_\mu(p,k)$, 
the incoming and outgoing lines represent $i G(p)$ and $ i G(p+k)$, 
and the broken line represents the external field 
carrying the four-momentum $k$. 

Next we show the proof~\cite{Ono} in the case of the heat current. 
If the interaction between electrons is local, 
the proof [N1,\ N4] can be carried out in real space as the above, 
only by replacing the charge current $j$ with the heat current $j^Q$, 
and we obtain 
\begin{equation}
\sum_{\mu=0}^3 {\partial \over \partial z_\mu} \Lambda_\mu^Q(x,y,z) = 
{\partial \over \partial x_0} G(x,y) \delta^4(z-x) 
+ 
{\partial \over \partial y_0} G(x,y) \delta^4(z-y), 
\label{del-t} 
\end{equation}
instead of (\ref{divLambda}). 
Since the interaction is local, 
the commutation relations 
$ [\rho^Q(z), \psi_\uparrow(x)] $ and $ [\rho^Q(z), \psi_\uparrow^\dag(y)] $ 
are equal to $ [K, \psi_\uparrow(x)] $ and $ [K, \psi_\uparrow^\dag(y)] $. 
Here $K = H - \zeta N$ with the Hamiltonian $H$, 
the total electron number $N$, and the chemical potential $\zeta$ and 
$K=\int d{\bf r} \rho^Q({\bf r})$. 
Thus the commutation relations lead to 
$ - i \partial \psi_\uparrow(x) / \partial x_0 $ and 
$ - i \partial \psi_\uparrow^\dag(y) / \partial y_0 $. 
This relation (\ref{del-t}) in real space 
is transformed into the Ward identity for the heat current, 
\begin{equation}
\sum_{\mu=0}^3 k_\mu \Gamma_\mu^Q(p,k) = 
p_0 G(p+k)^{-1} - (p_0+k_0) G(p)^{-1}. 
\label{Ward-Q} 
\end{equation}

If the interaction is non-local, 
we can obtain (\ref{Ward-Q}) only in the limit of ${\bf k}\rightarrow 0$. 
To prove this [N1,\ N2] we should use the Fourier variables as 
\begin{equation}
\sum_{\mu=0}^3 i k_\mu \Lambda_\mu^Q(p,k) = {\hat F} 
I_{{\bf p},{\bf k}}(x_0,y_0,z_0), 
\end{equation}
with 
\begin{equation}
{\hat F} \equiv 
\int d (x_0-y_0) e^{-ip_0(x_0-y_0)} \int d (z_0-x_0) e^{-ik_0(z_0-x_0)}, 
\end{equation}
where the integrand $I_{{\bf p},{\bf k}}(x_0,y_0,z_0)$ is given by 
\begin{equation}
\langle T 
\{ [\rho_{\bf k}^Q(z_0), a_{{\bf p}-{\bf k}}(x_0)] a_{\bf p}^\dag(y_0) \} 
\rangle \delta(z_0-x_0) 
+ 
\langle T 
\{ a_{{\bf p}-{\bf k}}(x_0) [\rho_{\bf k}^Q(z_0), a_{\bf p}^\dag(y_0)] \} 
\rangle \delta(z_0-y_0). 
\label{com-Q} 
\end{equation}
Here $\rho_{\bf k}^Q = 
\int d{\bf r} \exp(-i{\bf k}\cdot{\bf r}) \rho^Q({\bf r})$ and 
$a_{\bf p}$ and $a_{\bf p}^\dag$ are the Fourier transform of 
$\psi_\uparrow({\bf r})$ and $\psi_\uparrow^\dag({\bf r})$. 
In the limit of ${\bf k}\rightarrow 0$ [N1,\ N2] 
we can replace $\rho_{\bf k}^Q$ with $K$ in (\ref{com-Q}) 
and obtain (\ref{Ward-Q}). 

The above formulation at zero temperature 
is straightforwardly translated into that at finite temperature [N1,\ N4]. 

The Ward identities, 
(\ref{Ward-e}) for the charge current and (\ref{Ward-Q}) for the heat current, 
are consistent with the Jonson-Mahan formula~\cite{JM1} 
which is the exact relation between electric and thermal conductivities 
as discussed in [N1]. 

{\bf Diagrammatic proof.} The diagrammatic proof 
of the Ward identity (\ref{Ward-e}) for the charge current 
is a subject of a standard textbook~\cite{PS}. 
Most of the contributions to the current vertex cancel out 
and only \lq\lq end" contribution remains [N6]. 
The right-hand side of (\ref{Ward-e}) is the \lq\lq end" contribution. 
The cancelation in the case of the heat current is too complicated 
to show here, but the proof is given in [N6]. 
In the proof the Jonson-Mahan transmutation~\cite{JM2} plays a crucial role. 
It explains the way 
how the kinetic energy at the heat-current vertex for the free propagator 
is transmuted into the frequency for the full propagator. 
Thus the kinetic energy should be used 
at the heat-current vertex in the perturbatinal calculation. 
The frequency at the vertex 
appears after the renormalization of the interaction. 
This point is not recognized by most authors, 
so that their discussions become inconsistent. 
For an example, 
the violation of the Wiedemann-Franz law reported in~\cite{MF} 
is the consequence of the misuse of the frequency 
in the perturbational calculation. 
The perturbational calculation of the heat-current vertex for Cooper pairs 
reported in~\cite{Uss} should be criticized by the same reason. 

{\bf Local pairs.} The above algebraic proof for electrons is applicable 
to the case of local pairs [N1,\ N3,\ N4] 
only by replacing the annihilation and creation operators, 
$\psi_\uparrow(x)$ and $\psi_\uparrow^\dag(y)$, 
with those for pairs, $P(x)$ and $P^\dag(y)$. 
The particle-particle pair is given as 
$P^\dag(y)=\psi_\downarrow^\dag(y)\psi_\uparrow^\dag(y)$ 
where $\psi_\downarrow^\dag(y)$ is the creation operator 
of $\downarrow$-spin electron. 
The particle-hole pair is given as, for example, 
$P^\dag(y)=\psi_\uparrow^\dag(y)\psi_\uparrow(y)$ or 
$P^\dag(y)=\psi_\uparrow^\dag(y)\psi_\downarrow(y)$. 

In the case of charge current, 
instead of (\ref{com-ann}) and (\ref{com-cre}), 
the commutation relations for pairs are given by 
\begin{equation}
\big[ \rho(z), P(x) \big] \delta(z_0-x_0) 
= - e^* P(x) \delta^4(z-x), 
\end{equation}
and 
\begin{equation}
\big[ \rho(z), P^\dag(y) \big] \delta(z_0-y_0) 
= e^* P^\dag(y) \delta^4(z-y), 
\end{equation}
where $e^*$ is the charge of the pair. 
For the particle-particle pair $e^*=2e$ 
and $e^*=0$ for particle-hole pairs. 
Thus we obtain the Ward identity for pairs, 
\begin{equation}
\sum_{\mu=0}^3 k_\mu \Delta_\mu(q,k) = 
e^* D(q)^{-1}- e^*D(q+k)^{-1}, 
\label{Ward-pair-e} 
\end{equation}
where $\Delta_\mu$ is the charge-current vertex for pairs 
and $D(q)$ is the propagator for pairs with the four-momentum $q$. 
As depicted in Fig. 1 (right) 
$D(q)$ and the external field with four-momentum $k$ 
couple into $D(q+k)$ at $\Delta_\mu$. 

In the case of heat current we obtain the Ward identity for pairs, 
\begin{equation}
\sum_{\mu=0}^3 k_\mu \Delta_\mu^Q(q,k) = 
q_0 D(q+k)^{-1}- (q_0+k_0) D(q)^{-1}, 
\label{Ward-pair-Q} 
\end{equation}
by the same way as in the case of charge current. 

\color{red} 
Here we have discussed the current vertex for pairs. 
On the other hand, 
the internal structure of the vertex for electrons 
is discussed in terms of pair propagators in~\cite{HL}. 
\color{black} 

{\bf Extended  pairs.} To discuss the extended pairs [N2,\ N5] 
we introduce the center-of-mass coordinate ${\bf R}$ as 
\begin{equation}
P({\bf R}) = 
\int d {\bf r} \chi({\bf r}) 
\psi_\downarrow({\bf r}_1) \psi_\uparrow({\bf r}_2), 
\end{equation}
where ${\bf r}_1={\bf R}+{\bf r}/2$ 
and ${\bf r}_2={\bf R}-{\bf r}/2$. 
The Fourier transform of the particle-particle pair is given as 
\begin{equation}
P_{\bf q} = 
\sum_{\bf p} \chi({\bf p}) 
b_{-{\bf p}+{\bf q}/2} a_{{\bf p}+{\bf q}/2}, 
\end{equation}
where $b_{\bf p}$ is the Fourier transform of $\psi_\downarrow({\bf r})$. 
As seen in (\ref{com-Q}) 
it is essential for the derivation of the Ward identity 
to evaluate the equal-time commutation relation, 
$[\rho_{\bf k}, P_{\bf q}^\dag]$ for the charge current 
and 
$[\rho_{\bf k}^Q, P_{\bf q}^\dag]$ for the heat current. 
In the limit of ${\bf k}\rightarrow 0$ we obtain 
$[\rho_{\bf k}, P_{\bf q}^\dag] \rightarrow e^* P_{{\bf q}-{\bf k}}^\dag$ 
and 
$[\rho_{\bf k}^Q, P_{\bf q}^\dag] \rightarrow [K, P_{{\bf q}-{\bf k}}^\dag]$ 
so that the same Ward identities 
as (\ref{Ward-pair-e}) and (\ref{Ward-pair-Q}) result. 

{\bf Applications.} The Ward identity imposes a constraint 
on the vertex function and 
\color{red} can be a guide to a consistent theory\color{black}. 

In the study of superconducting fluctuation transport 
a relatively recent report~\cite{SSVG} claims the result not consistent 
with the time-dependent Ginzburg-Landau (TDGL) theory~\cite{USH,LV}. 
Since the TDGL theory is consistent with our Ward identities [N4] 
and obeys the conservation laws, 
it is concluded that such a claim violates the conservation laws. 
Our microscopic theory is consistent with the TDGL theory 
as the microscopic Fermi-liquid theory~\cite{AGD,YY} is consistent with 
the Boltzmann-transport theory. 

In the fluctuation-exchange (FLEX) approximation~\cite{Kon} 
discussing the transport anomaly 
in the normal state of high-$T_c$ superconductors, 
the Aslamazov-Larkin process of the particle-hole pair fluctuation 
vanishes [N3] in accordance with the Ward identity (\ref{Ward-pair-e}). 
On the other hand, if it vanishes, 
the FLEX approximation loses the consistency with the Fermi-liquid theory 
or the Boltzmann-transport theory. 
The inconsistency arises from the replacement 
of the renormalized interaction in the Fermi-liquid theory 
with the particle-hole fluctuation. 
Such a replacement violates the Pauli principle~\cite{BW,DHS,VT} 
essential for the degenerate Fermi systems. 
The correct microscopic treatment~\cite{AGD,YY} obeying the Pauli principle 
leads to the expected collision term in the Boltzmann equation. 

{\bf Concluding remarks.} In the discussion of the Ward identity 
the equal-time commutation relation plays the central role. 

In the case of the charge current 
it picks up the integrated charge of the object 
in the limit of vanishing external momentum. 
In this limit 
the wavelength of the electromagnetic field exceeds 
the size of the object 
so that the object can be treated as a point with its integrated charge 
in the discussion of the electromagnetic response. 
Thus it is concluded that charge-neutral pairs do not couple 
to electromagnetic field 
\color{red} as expected~\cite{KS}. 
\color{black} Namely, 
charge- and spin-density fluctuations do not carry charge. 
On the other hand, the particle-particle pair, 
the Cooper pair, carrying charge $2e$ 
couples to electromagnetic field. 

In the case of the heat current 
it picks up the energy of the object. 

\color{red}

Although only Ward identities for charge and heat currents 
are discussed in this Letter, 
other Ward identities are also actively discussed. 
For examples of recent developments, 
the spin current is discussed in~\cite{Fuj,GLHC}
and the sum rules are discussed in~\cite{YKM,GCH}.  

\color{black}

\vskip 10pt

The author is grateful to Kazumasa Miyake for illuminating discussions. 



\vskip 60pt 

\begin{figure}[htbp]
\begin{center}
\includegraphics[width=12.0cm]{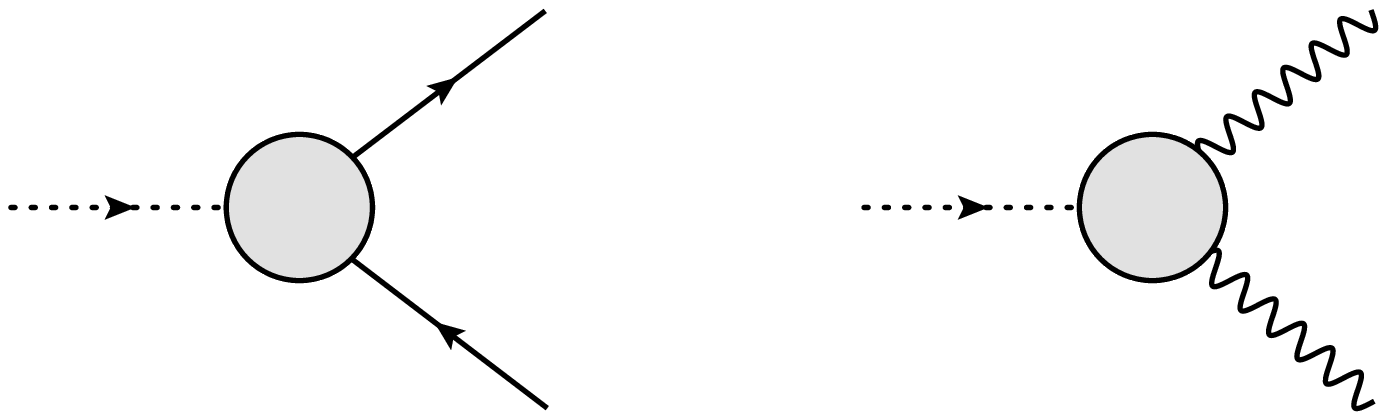}
\vskip 4mm
\caption{The current vertex for electrons (left) and pairs (right). }
\label{fig:vertex}
\end{center}
\end{figure}

\end{document}